\documentclass[a4paper]{jpconf}
\usepackage{graphicx}
\begin{document}
\title{Constraints on the Nuclear Symmetry Energy from Experiments, Theory and Observations}

\author{James M. Lattimer}

\address{Dept. of Physics \& Astronomy, Stony Brook University, Stony Brook, NY 11794-3800, USA}

\ead{james.lattimer@stonybrook.edu}

\begin{abstract}
Nuclear mass measurements and neutron matter theory tightly constrain the nuclear symmetry energy parameters $J$, $L$, $K_{sym}$ and $Q_{sym}$. Corroboration of these constraints on $J$ and $L$ can be found from measurements of the neutron skin thicknesses and dipole polarizabilities of neutron-rich nuclei. The experimental constraints on these parameters are compared with those obtained from consideration of astrophysical measurements of the neutron star radius, which we show is highly correlated with $L$. Attention is aimed at the recent PREX and CREX neutron skin measurements from Jefferson Lab, NICER neutron star radius measurements, and a new interpretation of the GW170817 tidal deformability measurement. We find joint satisfaction of PREX and CREX gives $J=32.2\pm1.7$ MeV and $L=52.9\pm13.2$ MeV, in excellent agreement with neutron matter predictions of $J=32.0\pm1.1$ MeV and $L=51.9\pm7.9$ MeV.
\end{abstract}

\section{Introduction}
The nuclear symmetry energy $S(n)$ is the focus of much recent activity
because it is the most direct link between nuclear physics and nuclear astrophysics~\cite{Steiner2005,LP07,Lattimer2021}.  Noting that the zero-temperature energy of uniform nuclear matter, $E(n,x)$, is a function only of the baryon density $n$ and the proton fraction $x$, we defined $S(n)=E_N(n)-E_{1/2}(n)$, the difference of the energies of pure neutron matter [PNM, $E_N(n)=E(n,x=0)$] and 
isospin symmetric nuclear matter [SNM, $E_{1/2}(n)=E(n,x=1/2)$].  
$E(n,x)$ can be expanded in powers of the neutron excess $1-2x$:
\begin{equation}
E(n,x)=E_{1/2}(n)+S_2(n)(1-2x)^2+S_3(n)(1-2x)^3 + \dots,
\label{eq:symm}
\end{equation}
but a common approximation is to retain only the quadratic term in
Equation~(\ref{eq:symm}) at every density, even for small $x$, so that
$S(n)\simeq S_2(n)$.  Chiral Lagrangian effective field theory $\chi$EFT expansions~\cite{Weinberg1967, Weinberg1968} for PNM, nuclear matter with various admixtures of protons, and SNM, indicate~\cite{Wellenhofer2016} that this approximation appears
valid for all $x$ for densities where $\chi$EFT is itself valid.  For matter with densities below the nuclear saturation density $n_s\simeq0.155\pm0.005$ fm$^{-3}$, such as that in nuclei, experimental evidence for significant higher-than-quadratic contributions is lacking, but this could be partly due to the near-symmetric character of nuclei.  

$S_2(n)$ can be Taylor-expanded around the nuclear saturation density $n_s\simeq0.155\pm0.005$ fm$^{-3}$, and the expansion coefficients are the symmetry energy $J=S_2(n_s)$, slope $L=3n_s(dS_2/dn)_{n_s}$, incompressibility
$K_{sym}=9n_s^2(d^2S_2/dn^2)_{n_s}$, and skewness $Q_{sym}=27n_s^3(d^3S_2/dn^3)_{n_s}$   parameters. 
If $S=S_2$, it follows that the PNM energy and pressure become
  \begin{equation}
  E_N(n_s)=E(n_s,0)=J-B;\qquad P_N(n_s)=P(n_s,0)=Ln_s/3,
  \label{eq:s1}\end{equation}
  where $B\equiv-E_{1/2}(n_s)\simeq16\pm0.5$ MeV is the bulk binding energy parameter of SNM.

We also introduce incompressibility and skewness parameters for SNM, $K_{1/2}$ and $Q_{1/2}$, and for PNM, $K_N$ and $Q_N$, respectively.
  $K_{1/2}\simeq230\pm20$ MeV has been deduced from giant monopole resonances~\cite{Agrawal2003,Todd-Rutel2005}, but there is little direct experimental evidence for values for $Q_{1/2}, K_N$ or $Q_N$. 
Nuclear experiments (measurements of binding energies, neutron skin thicknesses and electric dipole polarizabilites) and theorerical investigations of neutron matter, however, establish strong correlations between $J$ and $L$ which constrain their values.   

Of recent considerable interest are the parity-violating electron scattering neutron skin experiments of $^{208}$Pb  (PREX-I and PREX-II)~\cite{Adhikari2021} and $^{48}$Ca (CREX)~\cite{Adhikari2022}.  These measure the mean square difference of the neutron and proton radii using what is argued to be the most direct and least model-dependent technique to date~\cite{Thiel2019}. 
However, these experiments gave results that are barely compatible with each other; no conventional nuclear interaction can fit both measurements to the 68\% confidence level~\cite{Zhang2022,Reinhard2022}.  PREX, in particular, seems incompatible with previous neutron skin experiments and other expectations at greater than the $1\sigma$ level.  We attempt to identify the properties of nuclear interactions (that have already been fit to nuclear binding energies and charge radii) that best simultaneously satisfy both PREX and CREX measurements. We find ranges of $J$ and $L$ that not only agree with previous neutron skin results, other nuclear experiments, and neutron matter theory, but also astrophysical observations.  In comparison, Ref.~\cite{Zhang2022} concluded that joint PREX/CREX satisfaction leads to smaller central values of $J$ and $L$ with larger uncertainties than we find.

\section{The Link Between Nuclear Physics and Astrophysics}
The iconic connection between nuclear physics and astrophysics is the high degree of correlation between the pressure of neutron star matter (NSM) near $n_s$ and typical neutron star radii near $1.4M_\odot$, which   
was empirically established some time ago by Ref.~\cite{LP01} and refined by Ref.~\cite{Lattimer2013},
\begin{equation}
   R_{1.4}\simeq (9.51\pm0.49)(P_{NSM}/{\rm MeV~fm}^{-3})^{1/4}{\rm~km}.
\label{eq:r14p}\end{equation}
 NSM has a finite proton (electron) fraction determined by energy minimization with respect to $x$ at every density, leading to beta-equilibrium. 

The total energy per baryon of NSM in the quadratic approximation for Eq. (\ref{eq:symm}) is
\begin{equation}
    E_{NSM}=E_N-4xS(1-x)+{3\over4}\hbar c x(3\pi^2nx)^{1/3}.\label{eq:ensm}
\end{equation}
The last term is the electron contribution. Beta equilibrium requires $\partial E_{NSM}/\partial x=0$, giving
\begin{equation}
    x=\left(4S\over\hbar c\right)^3{(1-2x)^3\over3\pi^2n}.
\label{eq:xsv}
\end{equation}
At $n_s$, where only the leading order term  (proportional to $L$) contributes in a density expansion of the pressure about $n_s$, one finds to lowest order in $x$,
\begin{equation}
P_{NSM}(n_s)=\left(n^2{\partial E_{NSM}\over\partial n}\right)_{n_s}\simeq{Ln_s\over3}\left[1-\left({4J\over\hbar c}\right)^3{4-3J/L\over3\pi^2n_s}\right].
\label{eq:pnsm}\end{equation}
Thus the $P_{NSM}-R_{1.4}$ correlation is fundamentally dependent upon the nuclear symmetry energy.  The $J$ dependence is weak, but, in any event, $J$ and $L$ are highly correlated (see \S\ref{sec:correlations}). 
Thus, to lowest order, this represents primarily an $R_{1.4}-L$ relation.  However, inasmuch as the most sensitive density for the $P_{NSM}-R_{1.4}$ correlation is $1.5-2n_s$~\cite{Drischler2021a}, variations in the parameter $K_{sym}$ among interactions contribute to the uncertainty of the $R_{1.4}-L$ correlation, becoming especially important for large $L$.  Using available data for many published non-relativistic (Skyrme) and relativistic (RMF) interactions, as shown in Fig. \ref{fig:lr14}, the $R_{1.4}-L$ correlation implied by Eq. (\ref{eq:r14p}), including its uncertainties is confirmed (despite  Eq.~(\ref{eq:r14p}) having been  estimated using a small number of interactions).

\section{Correlations from Nuclear Binding Energies and Neutron Matter Theory\label{sec:correlations}}
Generally speaking, a specific technique (experiment or theory) is not able to fix individual symmetry energy parameters precisely, but Bayesian or least-squares fitting of the relevant data can form a confidence ellipse. 
In the case of the $J-L$ correlation, the confidence ellipse has a slope that depends on the density that the technique most effectively probes.  
for example, assume that the density-dependent symmetry energy $S$  has the simple form~\cite{Tsang2009}
\begin{equation}
S(u)=C_Ku^{2/3}+C_Vu^\gamma,\label{eq:s3}\end{equation}
where $u=n/n_s$ and $C_K\simeq12.5$ MeV and $C_V$ represent kinetic and potential contributions, respectively.
$\gamma$ and $C_V$ are parameters to be determined.  One therefore has
\begin{equation}
    C_V=J-C_K,\qquad L=2C_K+3\gamma(J-C_K),\quad \gamma={L-2C_K\over3(J-C_K)}.
\label{eq:jl}
\end{equation}
 To retain a good fit as $J$ and $L$ are varied around the $\chi^2$ minimum, $S(u)$ should remain invariant at its most effectively-probed density $u_e$~\cite{Lynch2022}.  This implies that \begin{equation}
    {\Delta J\over\Delta L}=-\left({\partial S(u)/\partial\gamma\over\partial S(u)/\partial J}\right)_{u_e}{\partial\gamma\over\partial L}=-{\ln u_e\over3}.
\label{eq:ue}\end{equation}
Therefore, a slope $\Delta L/\Delta J$ implies $S(u)$ is best determined at $u_e=\exp(-3\Delta J/\Delta L)$; the weaker the density-dependence of the symmetry energy, the closer $u_e$ is to 1.

\subsection{Nuclear Binding Energies}
The simple nuclear liquid drop model ~\cite{Myers1966}, consisting of volume, surface, and Coulomb energies
\begin{equation}
    E_{LD}(A,I)=\left[-B+JI^2\right]A+\left[E_S-S_SI^2\right]A^{2/3}+E_{Coul},
    \end{equation}
 where $E_S$ is the surface energy of symmetric nuclei and $S_S$ is the surface symmetry energy, results in a the strong $J-S_S$ correlation.
        Here $I=1-2Z/A$.  Additionally, one should consider shell and pairing energies, but these can be effectively eliminated by taking half the difference of the measured energies $E_{exp}(A,I)$ of nuclei having the same mass but values of $Z$ and $N$ each differing by two units.  This procedure leaves a residual Coulomb energy contribution $\Delta {\cal S}_C=-(6/5)(e^2A/R)(1-I)$. 
The optimum values of $J$ and $S_S$ can be found by minimizing
    \begin{equation}
        \chi^2=\sum_i^{\cal N}{[{\cal S}_{exp}(A_i,I_i)-{\cal S}(A_i,I_i)]^2\over{\cal N}},\quad  {\cal S}(A,I)\simeq AI^2\left(J-S_SA^{-1/3}\right)+\Delta {\cal S}_C
        \end{equation}
with respect to them.  ${\cal S}$ (${\cal S}_{exp}$) is the model (measured) symmetry energy of each of the ${\cal N}$ nuclides under consideration. The result is a confidence ellipse centered at $J_0$ and $S_{S0}$ with
\begin{equation}
    \sigma_J=\sqrt{\chi^{-1}_{JJ}},\quad \sigma_S=\sqrt{\chi^{-1}_{SS}},\quad {\Delta S_S\over\Delta J}=\cot\left[{1\over2}\tan^{-1}{2\chi_{JS}\over\chi_{SS}-\chi_{JJ}}\right],\quad  r={\chi_{JS}\over\sqrt{\chi_{JJ}\chi_{SS}}},
\end{equation}
where $\chi^{-1}$ is the inverse of  $\chi_{ij}=\partial^2\chi^2/\partial S_i\partial S_j$, $S_J=J$ and the correlation is $r$.   Since ${\cal S}$ is linear in $J$ and $S_S$, $\chi_{ij}$ depends only on $A_i$ and $I_i$ values, not on the optimized $J_0$ or $S_{S0}$.
Numerical values are obtained by using the set of 2336 nuclei from Ref.~\cite{Audi2003} with $N\ge40$ or $Z\ge40$: 
\begin{equation}
    \left[\chi_{JJ},\chi_{SJ},\chi_{SS}\right]={2\over{\cal N}}\sum_iI_i^4\left[A_i^2,-A_i^{5/3},A_i^{4/3}\right]\simeq[61.6,-10.7,1.87].
\end{equation}
 As a result, one finds $J_0=29$ MeV, $S_{S0}=50$ MeV, $\sigma_J=2.3$ MeV, $\sigma_S=13.2$ MeV, $\Delta S_S/\Delta J\simeq5.8$ and $r=0.997$, which represents a high degree of correlation. 

$S_S$ originates from an integration of $S(n)$ over the nuclear density profile $n(r)$ in a nucleus.  For non-relativistic Skyrme-like interactions, it can be shown~\cite{Steiner2005} that
\begin{equation}
    {S_S\over J}=\sqrt{Qn_s^3}\int_0^1\left({J\over S(u)}-1\right)\sqrt{u\over E(u,1/2)+B}du,
\end{equation}
where $Q$ is a gradient coefficient, determined by the Skyrme interaction parameters, that appears in the total Hamiltonian density of symmetric matter ${\cal H}_{1/2}={\cal H}_{B,1/2}+Q(\nabla n)^2$.
${\cal H}_{B,1/2}=nE_{1/2}(n)$ is the symmetric uniform matter bulk energy density.  For 240 Skyrme-like interactions~\cite{Dutra2012}, a high degree of linearity is found, and the liquid drop model then predicts 
\begin{equation}
    {S_S\over J}\simeq {0.0234L\over{\rm MeV}} + 0.764\pm0.130,\quad {\Delta L\over\Delta J}={1\over0.0234J_0}\left({\Delta S_S\over\Delta J}-{S_{S0}\over J_0}\right)\simeq6.0,
    \label{eq:ss}
\end{equation}
    meaning that binding energies constrain $S$ most effectively at the density $u_e\simeq0.61$.

\begin{figure}[h]
\vspace*{-1.cm}
\hspace*{-1.5cm}
\includegraphics[width=10.3 cm,angle=180]{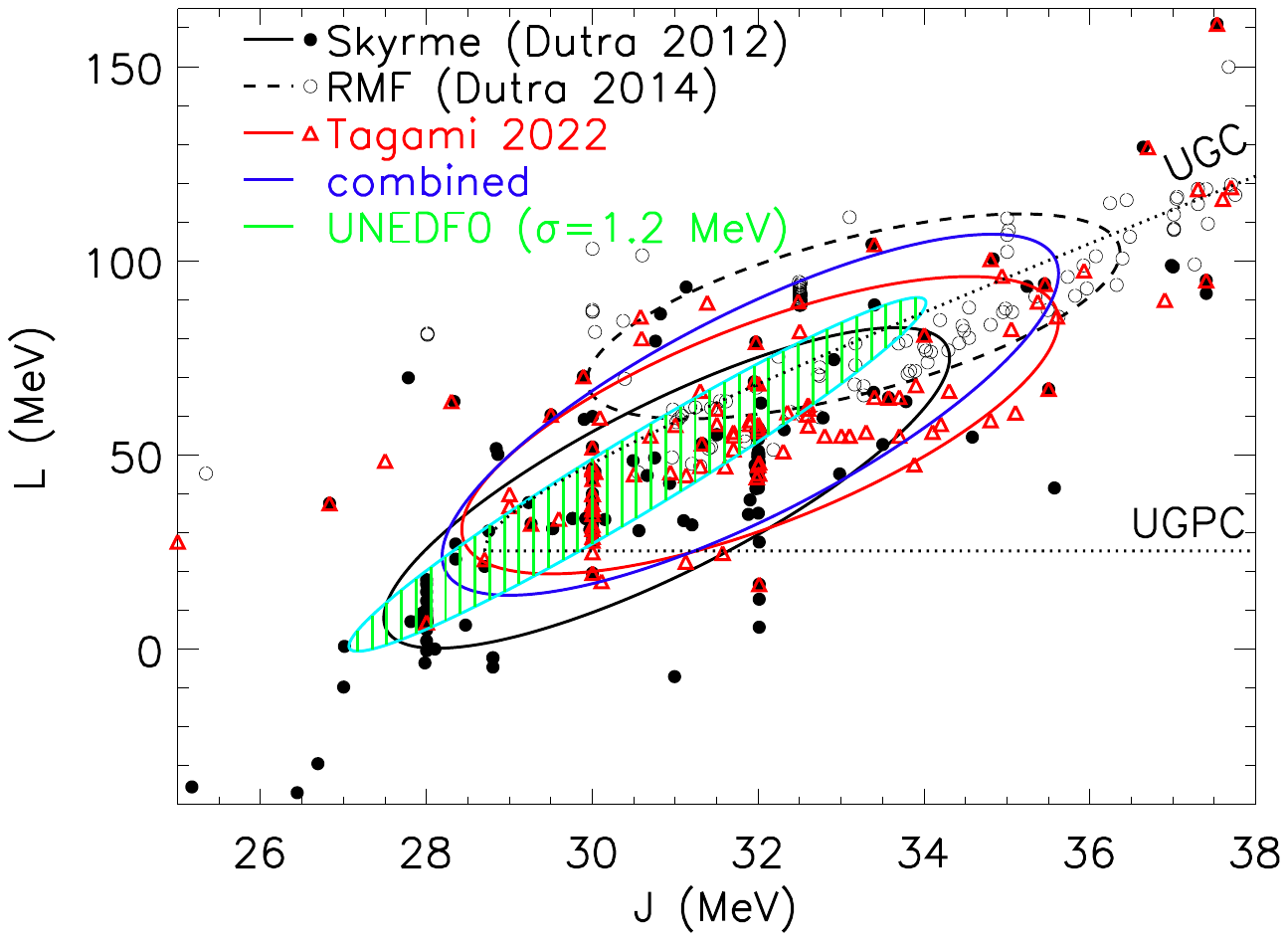}
\hspace*{-2.25cm}\includegraphics[width=10.3cm,angle=180]{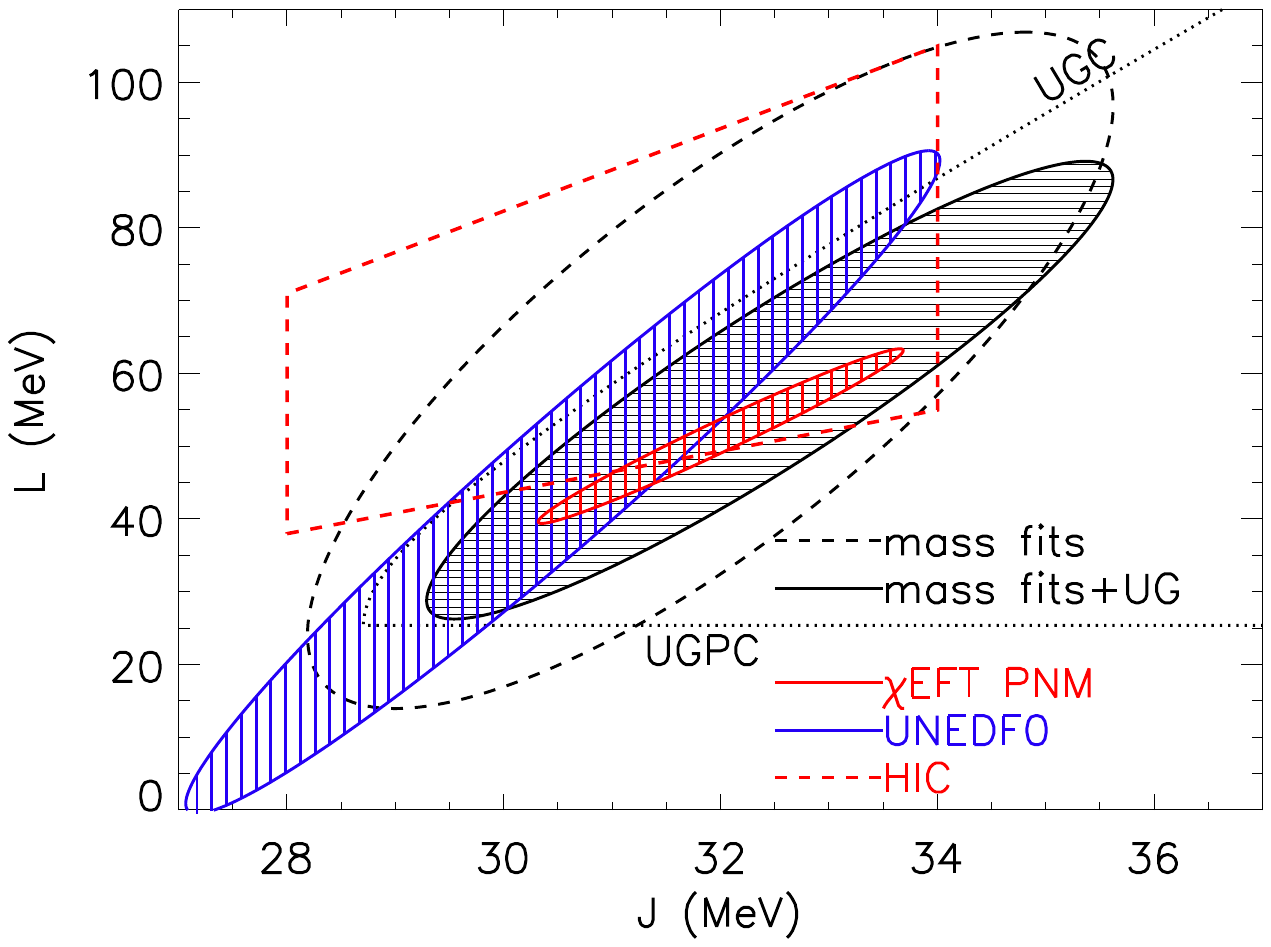}
\vspace*{-1cm}
\caption{Left: $J$ and $L$ data from individual Skyrme~\cite{Dutra2012} (black filled circles), relativistic mean field~\cite{Dutra2014} (black open circles) forces, both interaction types~\cite{Tagami2022} (red triangles), and all combined tabulated interactions; corresponding 68.3\% confidence ellipses are shown in black solid, black dashed, red and blue, respectively. 
Right:  Black correlation ellipses use model interaction data~\cite{Dutra2012,Dutra2014,Tagami2022}, with (solid)  and without (dashed) application of the UGC~\cite{Tews2017} and UGPC~\cite{Lattimer2023} boundaries shown as dotted curves in both panels.  The red confidence ellipse is from chiral EFT studies~\cite{Drischler2020} using PNM results with empirical saturation properties.  The red-dashed quadrilateral are limits determined from elliptic flows in heavy-ion collisions~\cite{Tsang2009}.   The green (blue) hatched confidence ellipse in the left (right) panel is taken from the UNEDF collaboration~\cite{Kortelainen2010} using $\sigma=1.2$ MeV. }
\label{fig:correl}
\end{figure}
The liquid droplet model~\cite{Myers1969}, for which 
$S(A,I)=AI^2J(1+S_SA^{-1/3}/J)^{-1}+\Delta {\cal S}_C$,
provides an improved
fit with $J_0=30.5$ MeV,  $S_{S0}=54.1$, 
a larger slope $\Delta L/\Delta J\simeq12$, and $u_e\simeq0.78$, but with a similar high degree of correlation.
The predicted liquid droplet $J-L$ correlation is confirmed from parameters in compilations of hundreds of non-relativistic interactions~\cite{Dutra2012}, relativistic interactions~\cite{Dutra2014}, and interactions of both types~\cite{Tagami2022} (all of which are fit to binding energies and selected other properties of nuclei to various degrees), as seen in Fig. \ref{fig:correl}.  These have, respectively, slopes of 12.1, 8.2 and 10.6.  In addition, the UNEDF collaboration~\cite{Kortelainen2010} explored the $J-L$ correlation in a systematic fashion  using a universal energy density functional fit to binding energies and charge radii of selected closed-shell nuclei.  Its optimized parameters are $J_0=30.5$ MeV and $L_0=45.1$ MeV, and its slope, 12.6 ($u_e=0.79$), is close to that obtained from the non-relativistic compilation of Ref.~\cite{Dutra2012}.  And choosing Ref.~\cite{Kortelainen2010}'s fiducial uncertainty parameter to be $\sigma\simeq1.2$ MeV yields an approximately equal uncertainty $\sigma_J$, as seen in Figure \ref{fig:correl}.
This is not surprising since the UNEDF universal energy density functional is non-relativistic.  However, the UNEDF $J-L$ correlation is much stronger than those found from compilations, possibly because the latter do not have the same strict calibrations involving charge radii.  

\subsection{Neutron Matter Theory}

A major recent advance in the understanding of nuclear matter has been made possible through the development of chiral effective field theory ($\chi$EFT)~\cite{Weinberg1967, Weinberg1968} which provides the only known framework allowing a systematic expansion of nuclear forces at low energies~\cite{Epelbaum2009,Machleidt2011,Hammer2020,Tews2020a} based on
the symmetries of quantum chromodynamics, the fundamental theory of the strong
interaction. In particular, $\chi$EFT allows one to derive systematic estimates of uncertainties
of thermodynamic quantities~\cite{Drischler2019,Leonhardt2020,Drischler2021a,Drischler2020b} for zero-temperature matter for densities up to $\sim2n_s$ with two- and three-nucleon interactions at the next-to-next-to-next-to-leading order (N3LO).

At this time, the $\chi$EFT calculations for PNM~\cite{Drischler2020} seem more reliable than for SNM.  In SNM, the predicted values of $n_s$ and $B$ are highly correlated, but their confidence ellipse does not come close to the empirical window obtained from non-relativistic and relativistic  compilations~\cite{Dutra2012,Dutra2014,Tagami2022} of interactions fit to nuclear binding energies~\cite{Lattimer2023}.   The failure $\chi$EFT to properly saturate is not surprising, considering that saturation in SNM emerges from a delicate cancellation sensitive to the short- and intermediate-range three-body interactions at next-to-next-to-leading order, in contrast to PNM where these interactions are Pauli-blocked~\cite{HLPS10}.

Fig. \ref{fig:correl} displays the $J-L$ confidence elllipse inferred using $\chi$EFT PNM results for the energy and pressure, including their standard deviations, evaluated at $n_s$, and coupled with $B$ values, both randomly chosen from within the empirical saturation window, and using Eq. (\ref{eq:s1}).  The correlation so determined is remarkably consistent with those obtained from binding energies, but has noticeably smaller uncertainties, a greater degree of significance, and a slightly larger slope.  Thus it appears that neutron matter calculations most senstitively probe the symmetry energy at a density $u_e\sim0.8$.

\subsection{The Unitary Gas Conjecture}
Ref.~\cite{Tews2017} proposed a constraint on the symmetry parameters arising from the conjecture that the PNM energy was 
greater, at all densities, than that of a unitary gas.  A unitary gas is an idealized theoretical collection of fermions interacting only via pairwise $s$-wave interactions with an infinite scattering length and a vanishing effective range.  The average particle separation in such a gas is the only length scale, so the energy of the unitary gas, $E_{UG}$, is proportional to the Fermi energy,
\begin{equation}
E_{UG}={3\hbar^2k_F^2\over10m_N}\xi_0\simeq12.7\left({n\over n_s}\right)^{2/3}=E_{UG,0}u^{2/3},
\label{eq:ug}\end{equation}
where $k_F=(3\pi^2n_s)^{2/3}$ is the Fermi wave number at the saturation density, $m_N$ is the neutron mass, the Bertsch parameter, which is experimentally measured~\cite{Ku2012,Zurn2013} to be $\xi_0\simeq0.37$, and $E_{UG,0}\simeq12.6$ MeV.  In reality, PNM at low densities has finite scattering length and range, each of which leads to energy increases.  In addition, three-body forces in PNM, which become increasingly important for $u>1$, are known to be repulsive, further increasing the energy.  

The Unitary Gas Conjecture (UGC) states that $E_N\ge E_{UG}$ at all densities.  If it is minimally satisfied $E_N(u_t)=E_{UG}(u_t)$ at some arbitrary density $u_t$, it is also required that~\cite{Tews2017}
\begin{equation}
\left({dE_N\over du}\right)_{u_t}=\left({dE_{UG}\over du}\right)_{u_t}
\label{eq:UGC}\end{equation}
 in order that the UGC remain satisfied at higher and lower densities. These two conditions automatically impose joint constraints on the parameters $J$ and $L$ and impose a minimum $J_{min}=B+E_{UG,0}$.  In addition, the Unitary Gas Pressure Conjecture, that the neutron matter pressure is always greater than the unitary gas pressure for densities larger than $n_s$, has been proposed~\cite{Lattimer2023}.  This sets a minimum $L_{min}=2E_{UG,0}$.  The resulting joint constraint on $J$ and $L$ is displayed in Figure \ref{fig:correl} and in subsequent figures; it is relatively insensitive to values higher-order symmetry parameters and ucertainties in $\xi_0, n_s$ and $B$~\cite{Tews2017}.  As anticipated, the UGC and UGPC are obeyed by the $\chi$EFT correlation.  Similarly, most non-relativistic interactions in the compilation of Ref.~\cite{Dutra2012} obey them while most relativistic forces in the compilation of Ref.~\cite{Dutra2014} do not. Correlations obtained from the interactions satisfying the UGC and UGPC have smaller uncertainties and become more consistent with the correlation from $\chi$EFT theory.

\section{Constraints from Neutron Skin Thicknesses}
That the neutron skin thickness of a neutron-rich nucleus depends strongly on $L$ and weakly on $J$ is well-known. The nuclides $^{48}$Ca and $^{208}$Pb are especially important because they are the only stable neutron-rich, closed-shell, nuclei and are thus spherical.   The liquid droplet model predicts 
the root mean square difference between the neutron and proton radii to be~\cite{Myers1969,Steiner2005}
\begin{equation}
r_{np}\simeq\sqrt{3\over5}{2r_o\over3}\left({1+{S_S\over JA^{1/3}}}\right)^{-1}\left[{S_S\over J}I-{3Ze^2\over140r_oJ}\left(1+{10\over3}{S_S\over JA^{1/3}}\right)\right],
\label{eq:tnp}\end{equation}
 where  $r_o=(4\pi n_s/3)^{-1/3}\simeq1.15$ fm.  The term proportional to $Z$ in the square brackets represents Coulomb polarization effects which reduce $r_{np}$.  The approximate neutron skin thicknesses of $^{48}$Ca and $^{208}$Pb are predicted by Eq. (\ref{eq:tnp}) as $r_{np}^{48}\simeq0.107$ fm and $r_{np}^{208}\simeq0.130$ fm.
 
 It is now possible to show from Eq. (\ref{eq:tnp}) that
\begin{equation}
    \left[{\Delta(S_S/J)\over\Delta J}\right]_{r_{np}}\simeq{Ze^2A^{1/3}\over70J_0^2}\left(r_{np}\sqrt{5/3}-{2r_o\over3}IA^{1/3}+{Ze^2\over21J_0}\right)^{-1}\simeq-0.020~(-0.0071),
\label{eq:dsdj}\end{equation}
and, using the first of Eq.(\ref{eq:ss}), that
\begin{equation}
    {\Delta L\over\Delta J}={\Delta L\over\Delta(S_S/J)}{\Delta(S_S/J)\over\Delta J}\simeq-0.84~(-0.30),
    \label{eq:dldj1}\end{equation}
  for $^{208}$Pb ($^{48}$Ca). 
    Note the negative values.  Therefore, the power of a series of precise neutron skin thickness measurements would be that they produce a confidence ellipse essentially orthogonal to the one from binding energies, helping to narrow the ranges of $J$ and $L$~\cite{Lattimer2013}.

\begin{figure}[h]
\vspace*{-1.3cm}
   \hspace{-1.5cm}  \includegraphics[width=10.5cm,angle=180]{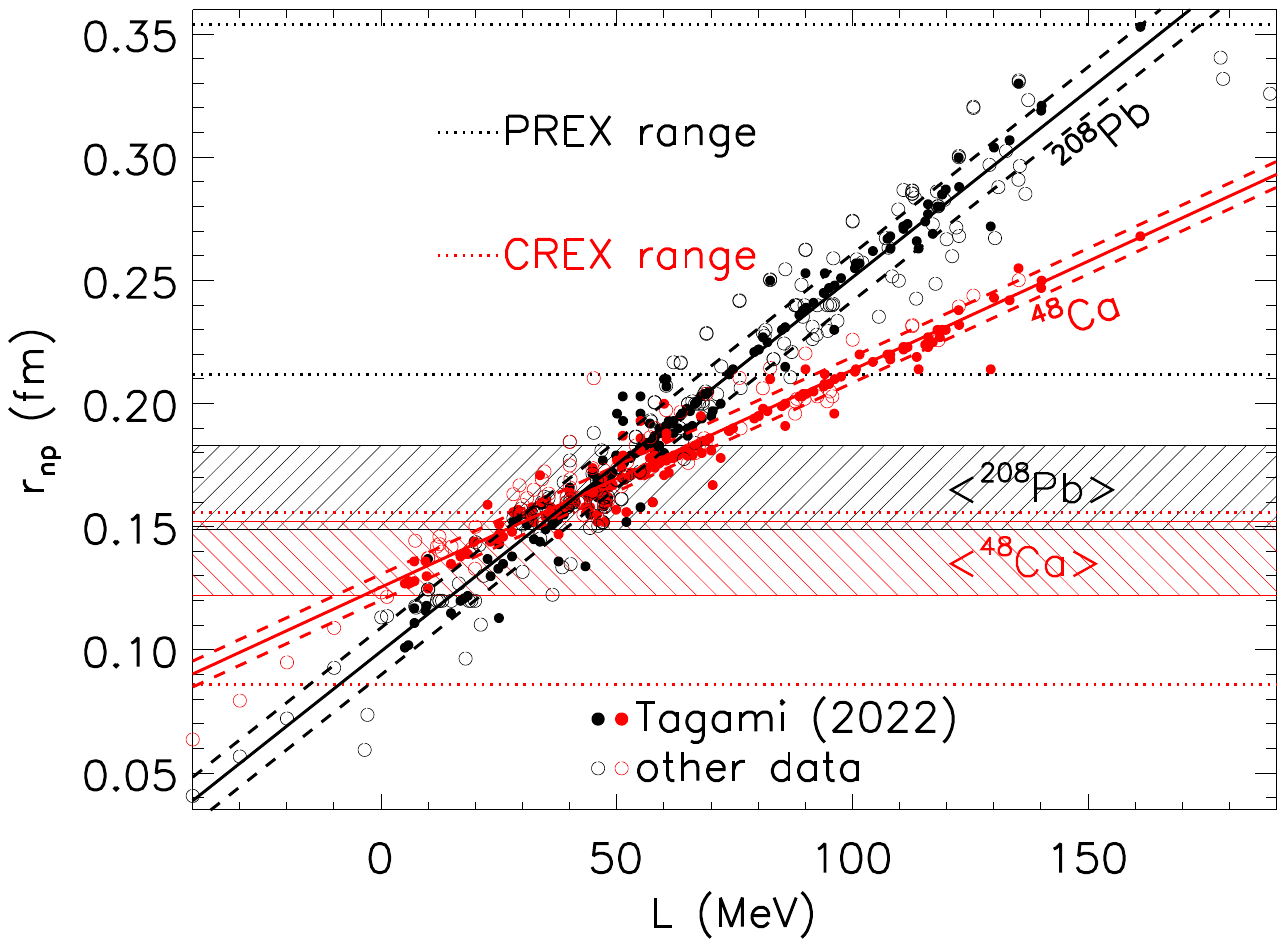}
   \hspace*{-2.5cm}\includegraphics[width=10.75cm,angle=180]{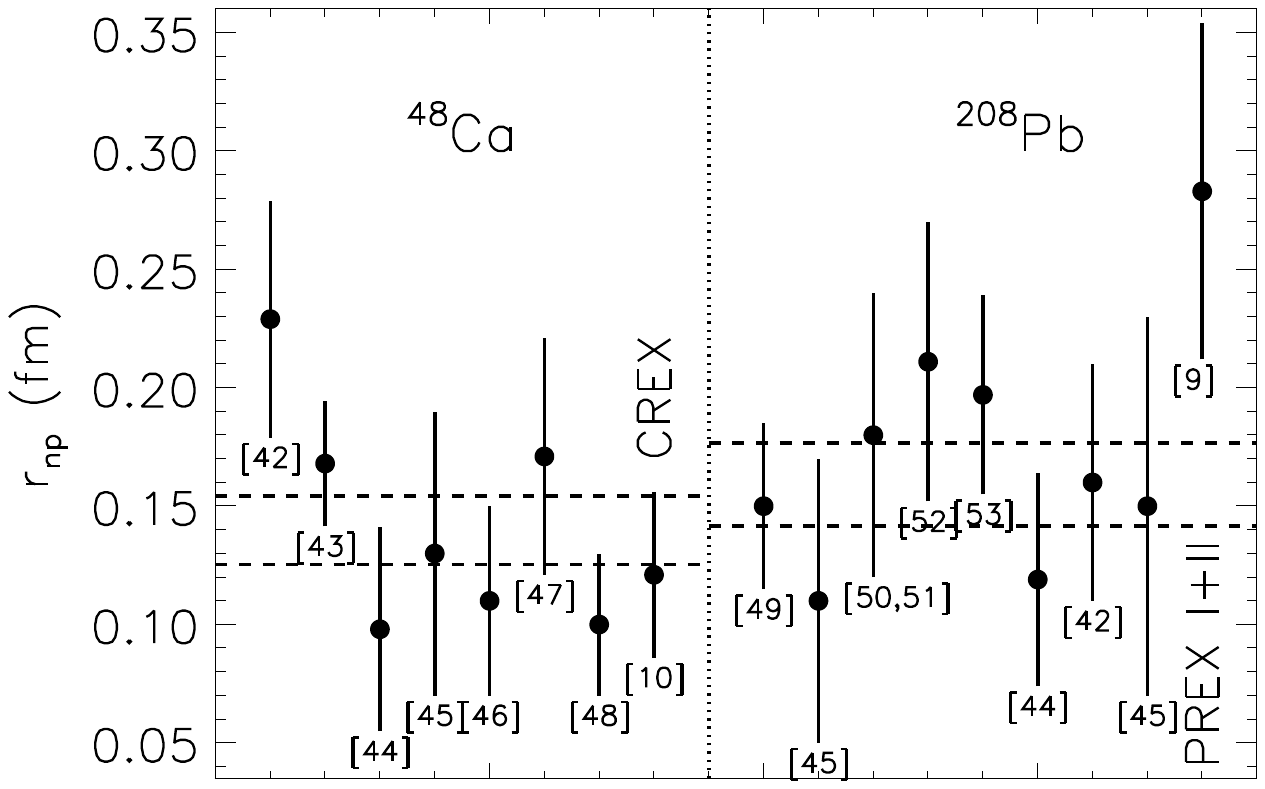}
  \vspace*{-1.1cm}  
  \caption{Left: Neutron skin thicknesses of $^{48}$Ca (red) and $^{208}$Pb (black) from interactions compiled in Ref.~\cite{Tagami2022} (filled circles) and Refs.~\cite{Brown2017,Adhikari2022,Horowitz2014,Piekarewicz2012} (open circles).  Means (1 standard deviations) of linear correlations are shown as solid (dashed) lines.  The horizontal shaded bands indicate the 1 standard deviation ranges of the averaged experimental results.  The dotted black (red) lines indicate the 1~standard deviation range of $r_{np}^{208}$ ($r_{np}^{48}$) from PREX I+II~\cite{Adhikari2021} (CREX~\cite{Adhikari2022}).
  Right: Neutron skin measurements~\cite{Ray1979,Zenhiro2018,Clark2003,Friedman2012,Gibbs1992,Gils1984,Shlomo1979,Adhikari2022,Tarbert2014,Klos2007,Brown2007,Zenhiro2010,Starodubsky1994,Adhikari2021} with 68\% confidence intervals and citations.  Horizontal dashed lines denote $\pm1$ standard deviations from the weighted means of experiments other than CREX or PREX I+II.  \label{fig:skin}}
\end{figure}

The dominant term in Eq. (\ref{eq:tnp}) is proportional to $IS_S/J$, suggesting an approximately linear relation between $r_{np}$ and $L$.    Linear relations for both $^{208}$Pb and $^{48}$Ca are indeed validated by examining recent compilations~\cite{Tagami2022,Adhikari2022,Brown2017,Horowitz2014,Piekarewicz2012
} using a multitude of both non-relativistic and relativistic interactions (Fig. \ref{fig:skin}). The compilation from Ref.~\cite{Tagami2022} is especially notable in containing results from 206 Skyrme-like and RMF forces, and the other compilations contribute more than 200 additional values. It is found, as seen in the left panel of Fig. \ref{fig:skin}, that
\begin{eqnarray}
    r_{np}^{48}&=&0.000882(L/{\rm MeV})+0.1255\pm0.0052{\rm~fm};\cr r_{np}^{208}&=&0.001518(L/{\rm MeV})+0.0996\pm0.0096{\rm~fm}.
\end{eqnarray}
   The mean values of the skin thicknesses from all models shown are nearly the same, $r_{np}^{208}=0.19\pm0.05$ fm and $r_{np}^{48}=0.18\pm0.03$ fm, but the two correlations have different slopes, largely because of differences in $I$.  Both mean values are within $\pm1\sigma$ of the respective mean experimental measurements (right panel of Fig. \ref{fig:skin}).
   For $^{208}$Pb, the error-weighted mean of all tabulated experimental measurements is $r_{np}^{208}=0.166\pm0.017$ fm.  The mean of historical measurements not including PREX is $r_{np}^{208}=0.159\pm0.017$ fm.  For $^{48}$Ca, the mean of all measurements is \mbox{$r_{np}^{48}=0.137\pm0.015$ fm}.  The average of historical measurements not including CREX is $r_{np}^{48}=0.140\pm0.017$ fm.  These results indicate an overall consistency exists between theory and experiment even though most of the models in the compilations were not explicitly fit to neutron skin values but rather to binding energy data. In other words, there is no reason to expect that either conventional interactions or modeling leads to large systematic uncertainties with respect to calculations of neutron skin thicknesses.

Combined, PREX I+II yields $r_{np}^{208}=0.283\pm0.071$ fm~\cite{Adhikari2021},  which Ref.~\cite{Reed2021} says translates to 68\% confidence ranges of $S_V=38.29\pm4.66$ MeV and $L=109.56\pm36.41$ MeV. Both values, and the measured value of $r_{np}^{208}$ itself, are considerably larger than from expectations from neutron matter and nuclear binding energies, and also from previous measurements, although overlapping with them at about the 90\% confidence level.  This indicates a tension with the current understanding of the equation of state (EOS).    In contrast, the measurement of the neutron skin of $^{48}$Ca using the same technique, $r_{np}^{48}=0.121\pm0.035$ fm~\cite{Adhikari2022}, is smaller than the average of earlier experimental measurements and expectations from nuclear binding energies and neutron matter theory.

Ref.~\cite{Zhang2022} performed a Bayesian analysis of the PREX and CREX results and found that the two experimental results are compatible at the 90\% confidence level.  Combining the data from the two experiments, they inferred $J=30.2^{+4.1}_{-3.0}$ MeV and $L=15.3^{+46.8}_{-41.5}$ MeV to 90\% confidence level, with mean values close to CREX expectations but far from PREX expectations.  Ref.~\cite{Reinhard2022} also performed a combined analysis, but conclude a simultaneous accurate description of the skins of $^{48}$Ca and $^{208}$Pb cannot be achieved with any non-relativistic or relativistic models that accommodate mass, charge radii and experimental dipole polarizabilities.

We follow two strategies for joint satisfaction of Ca-Pb measurements.  First, one could take the approach that the PREX and CREX experiments  qualitatively have fewer systematic uncertainties than other approaches, and only use those measurements.  Alternatively, an agnostic approach would be to instead consider the weighted means of all measurements.
 
\begin{figure}[h]
\vspace*{-1.cm}
\hspace*{-1.4cm}
\includegraphics[width=10.3cm,angle=180]{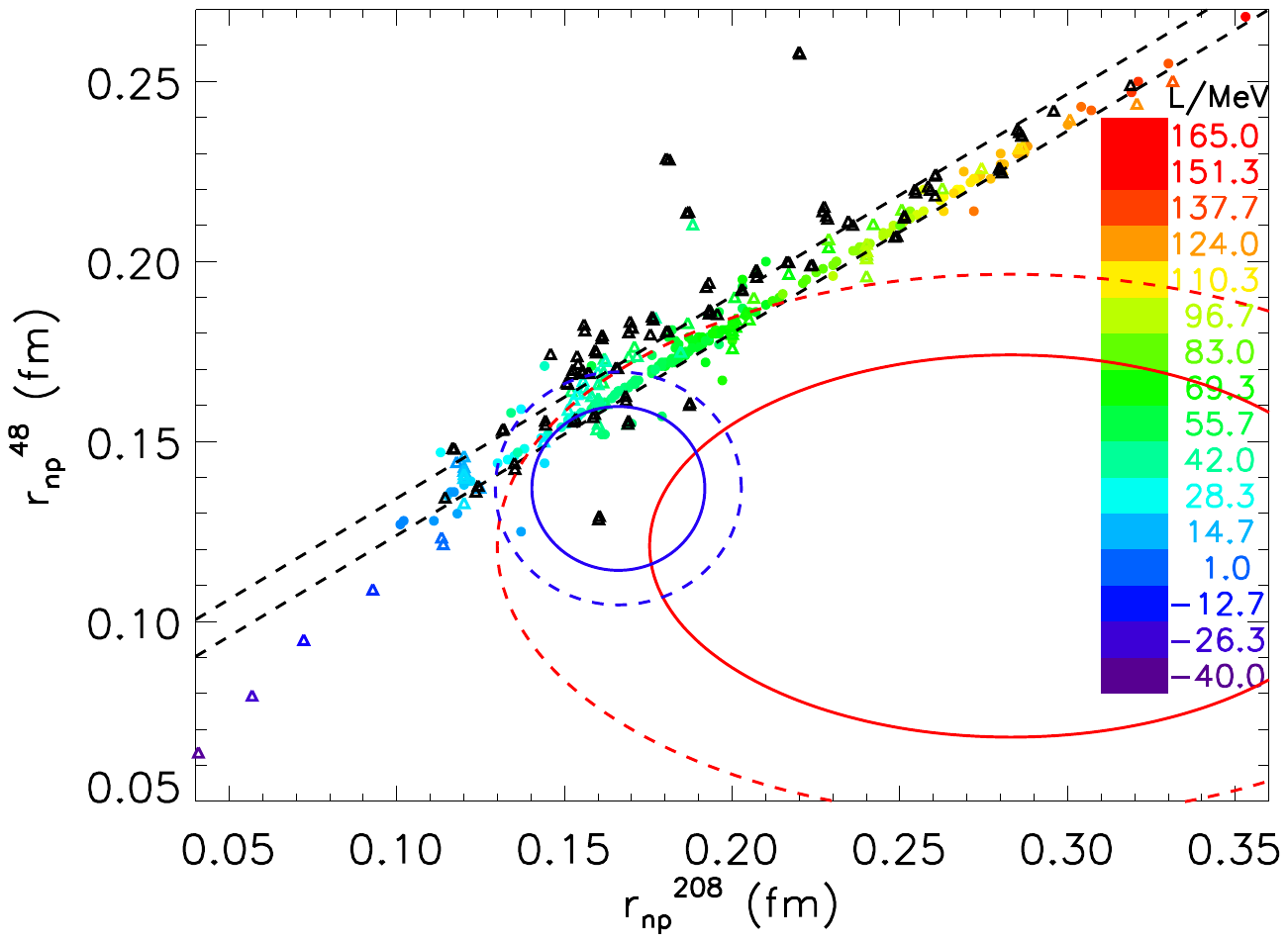}
\hspace*{-2.3cm}\includegraphics[width=10.3cm,angle=180]{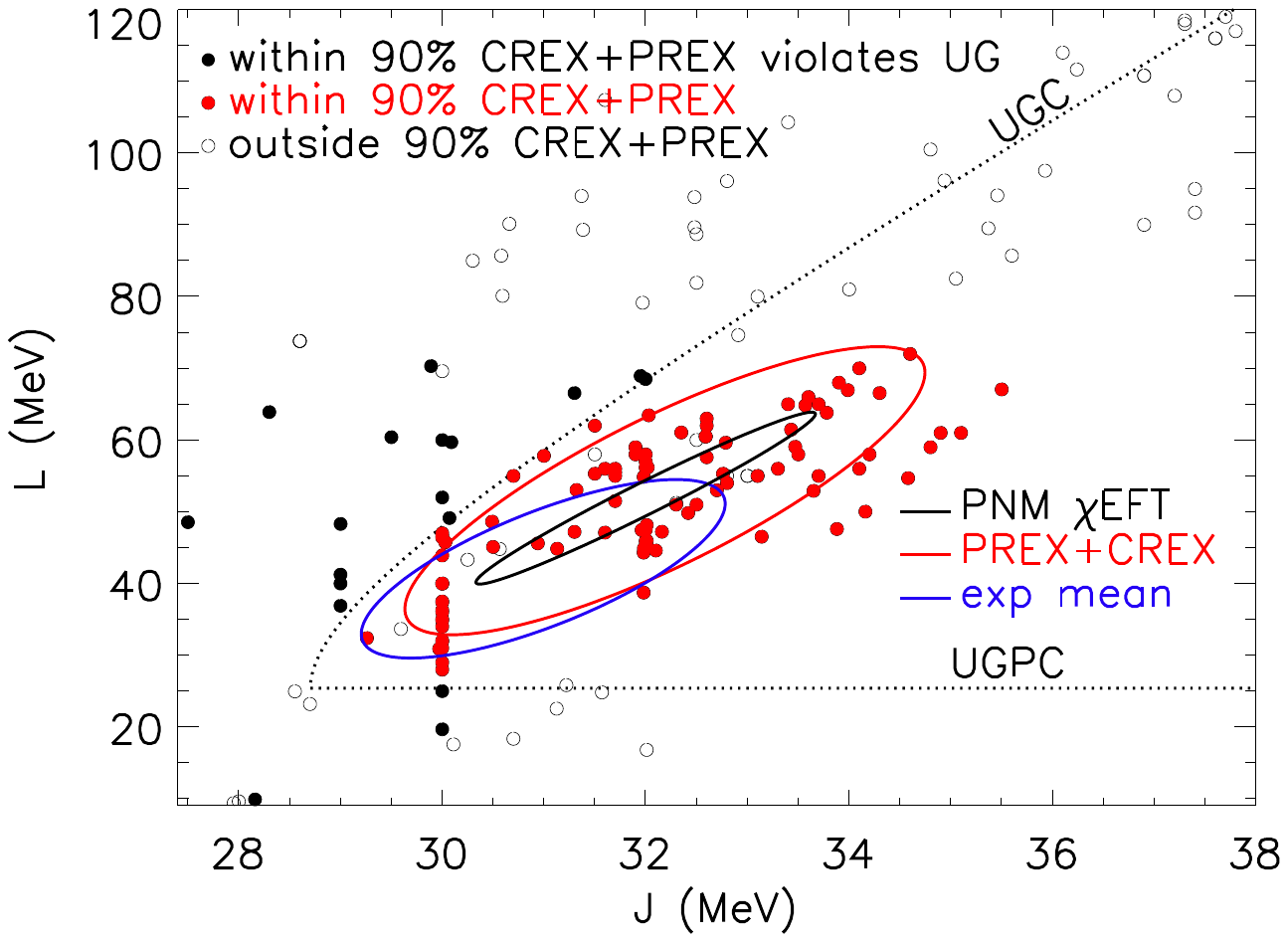}

\vspace*{-1cm}  \caption{Left: Neutron skin thicknesses of $^{48}$Ca and $^{208}$Pb from Ref.~\cite{Tagami2022} (filled circles) and Refs.~\cite{Brown2017, Adhikari2022,Horowitz2014,Piekarewicz2012} (triangles).  Color (black) indicate $L$ values where known (unknown).  Dashed lines show standard deviations from a linear $r_{np}^{48}-r_{np}^{208}$correlation.  The red (blue) confidence ellipses are from PREX I+II~\cite{Adhikari2021} and CREX~\cite{Adhikari2022} (mean of all experiments); solid (dashed) ellipses are 68\% (90\%) confidence. 
Right: Red and black filled (black open) circles show symmetry parameters $J-L$ jointly satisfying PREX/CREX to within (exceeding) 90\% confidence; filled black circles violate unitary gas constraints (dotted boundary).  Red (blue) confidence ellipses show models satisfying unitary gas constraints weighted by the two-dimensional Gaussian probability defined by PREX/CREX (red) and average (blue) experimental $r_{np}^{48}-r_{np}^{208}$ measurements and uncertainties.  The black confidence ellipse shows PNM $\chi$EFT results.
    \label{fig:capb}}
\end{figure}

As can be seen in Figure \ref{fig:capb}, the PREX I+II value for $r_{np}^{208}$ is too large and the CREX value for $r_{np}^{48}$ is too small to permit any of the reference interactions from the compilation of Refs.~\cite{Tagami2022,Adhikari2022,Brown2017,Horowitz2014,Piekarewicz2012} from satisfying both of them to within 68\% confidence.  The situation is different when considering the mean experimental results for $^{208}$Pb and $^{48}$Ca, with 4\% of the reference interactions simultaneously satisfying them to 68\% confidence. 

  A much larger number of interactions satisfy skin thickness measurements for both nuclei when considering 90\% confidence regions. About 40\% of the interactions from Ref.~\cite{Tagami2022} can simultaneously satisfy the UGC, UGPC, PREX I+II and CREX results, and these have 0 MeV $<L<72$ MeV.  
  Similarly, about 24\% of these interactions simultaneously lie within the 90\%  confidence region of the averages of all experiments, and also satisfy the UGC and UGPC, and these have 0 MeV $<L<58$ MeV.

The associated permitted region in $J-L$ space (right panel of Fig. \ref{fig:capb}) can be ascertained by weighting those interactions~\cite{Tagami2022} satisfying both unitary gas constraints, and which also have tabulated $J, L, r_{np}^{48}$ and $r_{np}^{208}$ values, 
by their probabilities given by a
two-dimensional Gaussian defined by the skin measurements and their uncertainties.   Interestingly, using CREX+PREX measurements to define the $r_{np}^{48}-r_{np}^{208}$  probabilities gives a confidence ellipse with $J=32.2\pm1.7$ MeV and $L=52.9\pm13.2$ MeV, in excellent agreement with neutron matter predictions of $J=32.0\pm1.1$ MeV and $L=51.9\pm7.9$ MeV.  Using the mean of other skin measurements gives somewhat smaller mean values of $J$ and $L$ by 1.2 MeV and 10.8 MeV, respectively. 

It is important to note that this internal consistency among neutron skin measurements, mass fitting and neutron matter theory, using either approach, suggests results are not particularly sensitive to whether relativistic or non-relativistic interactions are considered and so are relatively free of associated systematic uncertainties.  Partly, this is due to the moderate values of $L$ that are favored, eliminating most RMF interactions in compilations.

\section{Comparison with Astrophysical Observations}

\begin{figure}[h]
\vspace*{-1cm}\hspace*{-1.1cm}\includegraphics[width=10.2cm,angle=180]{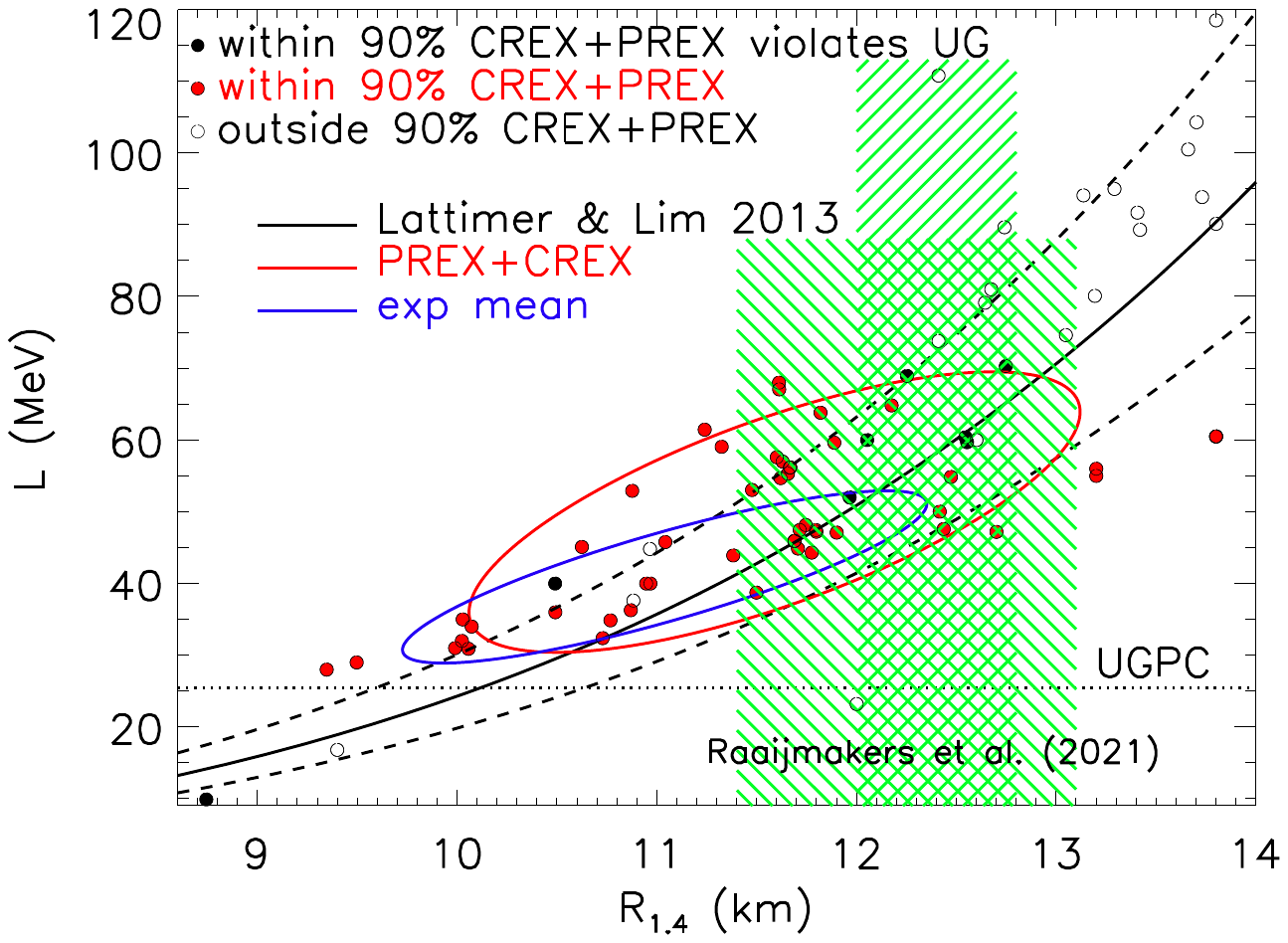}
\hspace*{-2.1cm}\includegraphics[width=10.2cm,angle=180]{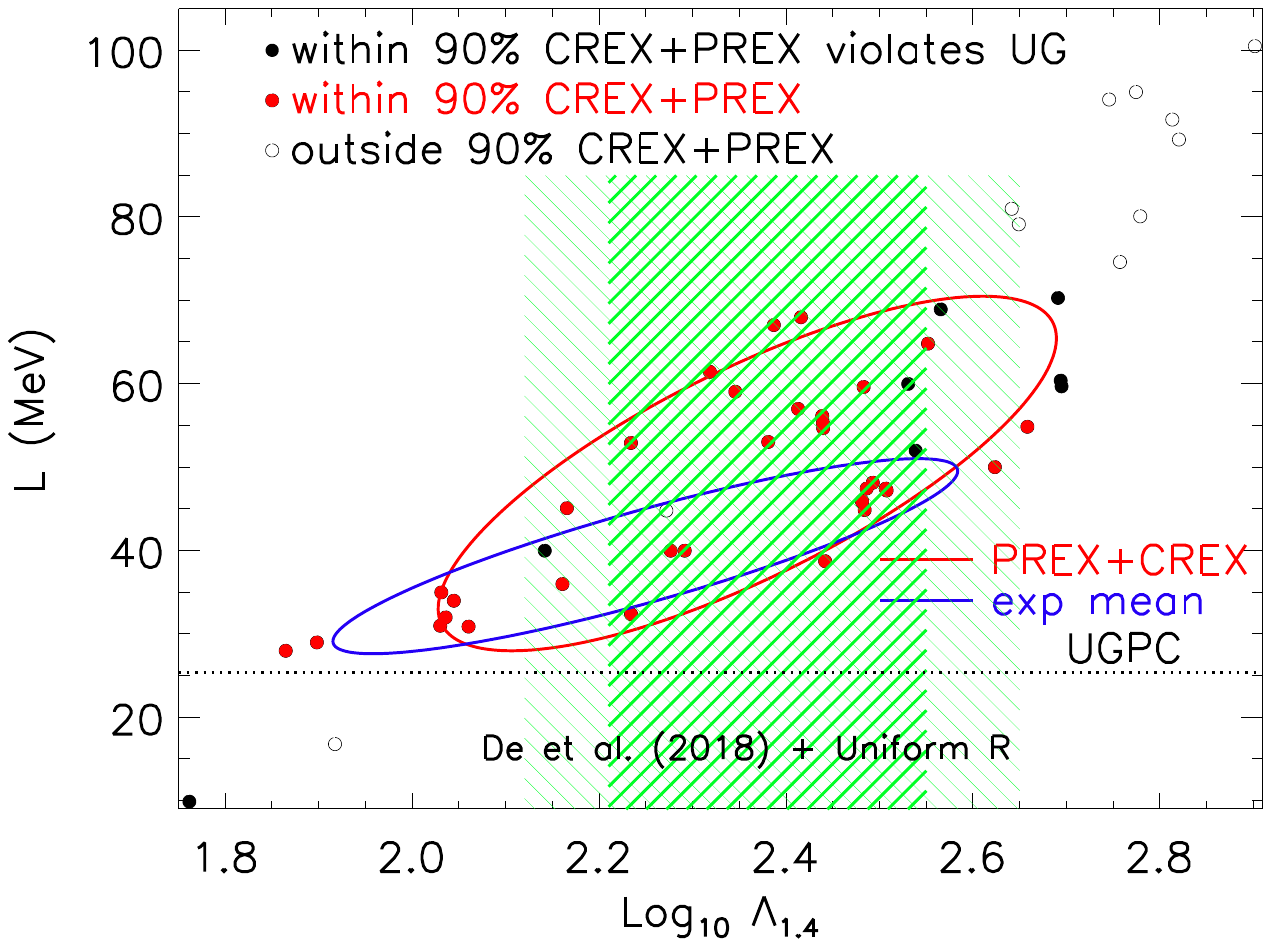}
\vspace*{-0.8cm}
\caption{The same as the right panel of Fig. \ref{fig:capb} except showing $R_{1.4}$ versus $L$ (\textbf{left panel}) and $\Lambda_{1.4}$ versus $L$ (\textbf{right panel}) for subsets of forces from Refs.~\cite{Tagami2022,Adhikari2022,Brown2017,Horowitz2014,Piekarewicz2012}.  Black solid and dashed curves in the left panel show the $R_{1.4}-L$ correlation and standard deviations derived as in the text. The shaded green bands are 68\% and 95\% confidence intervals from a joint analysis of GW170817 and PSR J0030+0451 and PSR J0740+6620 by Ref.~\cite{Raaijmakers2021} (\textbf{left panel}) and from GW170817 Bayesian analyses posteriors~\cite{De2018}, corrected for $\Lambda$ priors chosen so as to reflect uniform $R$ priors (\textbf{right panel}).}
\label{fig:lr14}\end{figure}

 One can predict validity ranges for $R_{1.4}$ and the corresponding tidal deformability $\Lambda_{1.4}$ based on neutron skin measurements in the same way as for the $J-L$ predictions shown in the right panel of Figure \ref{fig:capb}. Weighting the predicted values of $R_{1.4}$ and $\Lambda_{1.4}$ computed from compilations of models satisfying both unitary gas constraints that are weighted by their double Gaussian probabilities from PREX and CREX in $r_{np}^{48}$--$r_{np}^{208}$ space, one finds $R_{1.4}=11.6\pm1.0$ km and  $\Lambda_{1.4}=228^{+148}_{-90}$ to 68\% confidence using parity-violating neutron skin measurements (Figure \ref{fig:lr14}).    
In comparison, employing double Gaussian probabilities from the weighted average of other experiments yields $R_{1.4}=11.0\pm0.9$ km and  $\Lambda_{1.4}=177^{+117}_{-70}$. 
These limits have uncertainties comparable to inferences from astronomical measurements from NICER~\cite{Raaijmakers2021} of PSR J0030+0451 and PSR J0740+6620 and from the LIGO/Virgo observation of the binary NS (BNS) merger GW170817~\cite{De2018,Abbott2019}, but generally are very consistent with them.

\ack
Thanks are due to J. Piekarewicz, W. Nazarewicz and B. Tsang for providing data and, together with T. Zhao, C. J. Horowitz and K. Kumar, helpful discussions.  This work was initiated through the NSF-funded Physics Frontier Center Network for Neutrinos, Nuclear Astrophysics, and Symmetries (N3AS) and was supported by the U.S.~DOE grant
DE-AC02-87ER40317.

\end{document}